**trec: An R package for trend estimation and classification to support integrated ecosystem assessment of the marine ecosystem and environmental factors**

Hiroko Kato Solvang
Institute of Marine Research, Nordnes, N-5817, Bergen, Norway
(https://orcid.org/0000-0002-0330-4670)
hirokos@hi.no

Mineaki Ohishi
Hiroshima University, Naka-ku, Hiroshima, Japan
(https://orcid.org/0000-0001-8727-3631)
mineaki-ohishi@hiroshima-u.ac.jp

**Abstract**

Solvang and Planque (2020) provided a trend estimation and classification (TREC) approach to estimating dominant common trends among multivariate time series observations. This approach was developed to improve communication among stakeholders like marine managers, industry representatives, non-governmental organizations, and governmental agencies as they investigate the common tendencies between a biological community in a marine ecosystem and the local environmental factors. The entire calculation procedure was originally implemented using MATLAB (ver.R2018b). In this paper, we present R package **trec**, which was motivated by the requests of readers of the Journal of Marine Science, published by the International Council for the Exploration of the Sea. The tasks of trend estimation and classification in the original program have been revised, and new features include an automatic *icon* assignment algorithm using a multinomial logistic discriminator. Implementation of this package involves three partitions corresponding to TREC1) estimating trends from observed time series data; TREC2) classifying two/three rough patterns; and TREC3) generating a table summarizing categories of common configurations and the automatic *icon* assignments to them.

**Github repository**: https://github.com/ohishim/trec/

# Introduction

Integrated ecosystem assessment (IEA) is one approach to organizing scientific information at multiple scales and across sectors to support ecosystem-based fisheries management (EBFM) (Levin et al. 2009). IEA results can reflect various aspects of an ecosystem beyond dynamics, status, and future risks. In particular, identifying common trends can be useful as a diagnostic tool to reveal past changes and to explore the relationships among biological communities, as well as between these communities and environmental conditions. For such investigation, trend estimation and classification (TREC) was proposed in Solvang and Planque (2020).

TREC involves tasks for trend estimation of the observed time series data and for classification of the estimated trends into common trend groups. In the trend estimation task, the original TREC used two different kinds of parametric trend models: a polynomial regression model and a stochastic difference equation model. The trend estimated by the polynomial regression model presents a simpler trend configuration than that estimated by the stochastic trend model using a difference equation model. Solvang and Planque (2020) reported that the simpler patterns could be more robustly classified to common trend pattern groups than when using variable trend patterns by the stochastic difference equation model. Therefore, **trec** in R (R Core Team, 2016) adopts the usage of the simple polynomial regression model, which aims to achieve common trend classification.

In the trend classification task, two-category discrimination is first applied to roughly divide trends into three groups representing configurations for upward, flat, and downward. In the original procedure, a two-categorical discriminant function is calculated, and hierarchical clustering is applied by the discriminant function according to Solvang et al. (2008). The **trec** in R includes a procedure to check the outputs by the discriminant function with/without hierarchical clustering.

If it is necessary to classify trends into groups for more concrete common patterns, multiple-category discrimination is secondarily applied to the target trends, which the users can define in the estimated trends. Each target trend is assigned as a predefined reference, called an *icon*, which is set as an easily accessible form that can be used to serve the needs among stakeholders. In the original MATLAB code provided by Solvang and Planque (2020), the assignment to the multiple-category groups was done manually by the user. The **trec** in R has been developed to automatic assign *icon*s with a multinomial logistic regression model, which is an extended model for handling a response variable having more than two categories. Such a multinomial logistic regression model can be applied to multiple-discriminant analysis (e.g., Jeune et al., 2018; Mirzaei, 2019; Amri et al., 2020).

The **trec** in R is interactively implemented by the user in three procedures: TREC1, TREC2, and TREC3. In this article, we first introduce the methodological background in the next section and then give practical instructions for the **trec** package in the section 'Using **trec**'.

# Methodological background

## Trend model

The observation model of a time series is given by

$$y(n) = t(n) + u(n), \quad n = 1, \cdots, N$$

where $t(n)$ is the trend component and $u(n)$ is the residual component at time step $n$. In **trec**, the trend component is modelled by the following simple polynomial regression model, and the residual is assumed to be Gaussian white noise with mean zero and variance $\sigma^2$:

$$t(n) = \beta_0 + \beta_1 n + \cdots + \beta_{p-1} n^{p-1}$$

with an unknown vector $\boldsymbol{\beta} \equiv (\beta_0, \beta_1, \cdots, \beta_{p-1})' \in \mathbb{R}^p$. The apostrophe symbol in vector $\boldsymbol{\beta}$ denotes transposition. The least squares estimator for $\boldsymbol{\beta}$ is given by

$$\hat{\boldsymbol{\beta}}_{LSE} \equiv (\mathbf{Z}'\mathbf{Z})^{-1} \mathbf{Z}'\mathbf{Y}$$

where

$$\mathbf{Z} = \begin{pmatrix} 1, & 1, & 1^2, & \cdots, & 1^{p-1} \\ 1, & 2, & 2^2, & \cdots, & 2^{p-1} \\ & & \vdots & & \\ 1, & N, & N^2, & \cdots, & N^{p-1} \end{pmatrix} \text{ and } \mathbf{Y} \equiv (y(1), y(2), \cdots, y(N))'.$$

The trend component is estimated by

$$\hat{t}(n) \equiv (1, n, n^2, \cdots, n^{p-1}) \hat{\boldsymbol{\beta}}_{LSE},$$

which is defined on all time steps $\{1, 2, \cdots, N\}$. The estimated residual's variance obtained by the least squares prediction is given by

$$\hat{\sigma}^2 = \frac{1}{N} \sum_{n=1}^{N} \{y(n) - \hat{t}(n)\}^2.$$

Using the variance, the maximum log-likelihood of the polynomial trend model for this is given by

$$l(\hat{\boldsymbol{\beta}}, \hat{\sigma}^2) = -\frac{N}{2} (\log 2\pi + 1 + \log \hat{\sigma}^2).$$

Denoting the number of parameters by $c$, the Akaike Information Criterion (AIC) (Akaike, 1974) value of the model, given the estimate of optimum orders for the polynomial trend model, is

$$\text{AIC}(c) = -2l(\hat{\boldsymbol{\beta}}, \hat{\sigma}^2) + 2c.$$

**Trend classification**

For the trend estimation, we apply the discriminant analysis proposed by Solvang et al. (2008). First, we consider the following two-category discrimination:

$$\Pi_1: \text{ model (1) with trend function } T_1(n)$$
$$\Pi_2: \text{ model (1) with trend function } T_2(n)$$

Suppose that we observe new data and that the trend is estimated as $\hat{t}(n)$, which is assumed to belong to $\Pi_1$ or $\Pi_2$. In practical use, we fix two reference trends $\hat{T}_1(n)$ and $\hat{T}_2(n)$, which are estimates of $T_1(n)$ and $T_2(n)$. We define discriminant function $D \equiv L(\hat{t}:T_2) - L(\hat{t}:T_1)$ for use as a discriminant score (Solvang et al., 2008), where $L(\hat{t}:T_j) \equiv \sum_{n=1}^{N} \{T_j(n) - \hat{t}(n)\}^2$, $j = 1, 2$. It is assumed that in practice $\hat{T}_1(n)$ and $\hat{T}_2(n)$ represent different (opposite) shapes, namely increasing and decreasing. In **trec**, we use an artificial target trend set composed of a pair of trends with slopes having completely opposite shapes as default. The different shapes result in a large distance between the two categories. Then, the discriminant function for $m$ dimensional observation $y^{(i)}(n)$, $i = 1, 2, \cdots, m$ is defined by

$$\hat{D}_j = \sum_{n=1}^{N} \{\hat{T}_2(n) - \hat{t}_j(n)\}^2 - \{\hat{T}_1(n) - \hat{t}_j(n)\}^2$$

If $\hat{D}_j > 0$, category $\Pi_1$ is chosen, otherwise, category $\Pi_2$ is chosen. By applying hierarchical clustering with centroid linkage, we can classify groups according to similar $\hat{D}_j$ e.g. upward, downward and flat, including convex or concave shapes (Solvang et al., 2008).

Second, if it is necessary to classify trends into groups of more concrete patterns from the three rough-configuration groups, this two-category classification can be easily extended to more precise multiple-category discrimination.

While the original TREC extended multiple-category discrimination in general, the **trec** in R conducts multiple-category discrimination for each rough-configuration group $j$, e.g. upward and downward for $j = 2$, as below:

$$\Pi_{i,j} : \text{model (1) with trend function } T_i(n), \quad i = 1, 2, \cdots, m_j,$$

where $m_j$ is the maximum number of target trends set in group $j$. We provide a divergence measurement set for each rough-configuration group given by

$$\zeta_j = \{L(\hat{t} : T_1), L(\hat{t} : T_2), \cdots, L(\hat{t} : T_{k_j})\}.$$

The classification rule is defined as the requirement that the estimated trend $\hat{t}$ belongs to $\Pi_{l,j}$ to satisfy

$$L(\hat{t} : T_{l,j}) = \min(\zeta_j).$$

In practice, the reference trend $T_{i,j}$ should be predefined by the user into the three configurations groups obtained by the two-category discriminants. Finally, the reference trend $T_{i,j}$ is assigned as a general reference, called an *icon*, which we expect to work as an easily accessible form that can serve the needs of stakeholders. The entire numerical procedure and the {\it icon}s we predefined are summarized in Figure 1.

**Assignment of icons to common trend groups**

In the **trec** package, variables of multivariate time series data are classified into two or three groups, which are the upward, downward, and flat groups. Then, we consider the assignment of *icon*s in Figure 1 to each group. Specifically, for the variables of the upward group, one among *icon*s 3, 4, 5, and 10 is assigned, for the variables of the downward group, one among *icon*s 2, 7, 8, and 10 is assigned, and for the variables of the flat group, one among *icon*s 1, 6, 9, and 10 is assigned, where *icon* 10 is "?", which means none of *icon*s $1, \cdots, 9$ is suitable for a variable. We adopt a multinomial logistic discriminator for the assignment. That is, we estimate unknown parameters for a multinomial logistic model from a training dataset, which we use for the assignment, where the training dataset is made from the time series data compiled by the ICES integrated assessment working groups for the Barents Sea (WGIBAR; ICES, 2020) and the Norwegian Sea (WGINOR; ICES, 2020). Hence, a discriminator is constructed for each group. That is, for example, the discriminator for the upward group is made from the training dataset having only upward variables, and this discriminator outputs one of the *icon*s 3, 4, 5, or 10. Note that the training datasets for the upward, downward, and flat groups have 257, 276, and 137 variables, respectively. In the following, we describe how the discriminator for the upward group is constructed. Those for the remaining two groups can be constructed in the same way.

For the training dataset of the upward group, let $a_i \in \{1,2,3,4\}, i=1,\cdots,m_1$ be indexes expressing the *icon* of the $i$ th variable, where the indexes 1, 2, 3, and 4 correspond to *icon*s 3, 4, 5, and 10, respectively, and $m_1$ is the number of variables of the training dataset. The multinomial logistic discriminator evaluates the probability of each choice of *icon* (1, 2, 3, or 4) and outputs that choice with the highest probability. Let $\mathbf{A}_i \equiv (a_{i,1},\cdots,a_{i,4})'$ be a four-dimensional vector defined by

$$a_{i,j} \equiv \begin{cases} 1 & (a_i = j) \\ 0 & (\text{otherwise}) \end{cases}, \quad j=1,\cdots,4.$$

Then, we consider the multinomial logistic model that $\mathbf{A}_i$ is distributed according to the multinomial distribution with the number of event 1 and the cell probability vector $(\pi_{i,1}(\theta),\cdots,\pi_{i,4}(\theta))'$, given by

$$\pi_{i,j}(\boldsymbol{\theta}) \equiv \begin{cases} \dfrac{\exp(\mathbf{X}_i'\boldsymbol{\theta}_j)}{1+\sum_{l=1}^{3}\exp(\mathbf{X}_i'\boldsymbol{\theta}_l)} & (j=1,2,3) \\ \dfrac{1}{1+\sum_{l=1}^{3}\exp(\mathbf{X}_i'\boldsymbol{\theta}_l)} & (j=4) \end{cases}, \quad i=1,\cdots,m_1,$$

where $\mathbf{X}_i$ is a seven-dimensional vector obtained from the training dataset (defined later) and $\boldsymbol{\theta} \equiv (\boldsymbol{\theta}_1',\cdots,\boldsymbol{\theta}_4')$ of which $\boldsymbol{\theta}_j$, $j=1,\cdots,4$ are seven-dimensional vectors of unknown parameters. By estimating unknown parameter vector $\boldsymbol{\theta}$, the multinomial logistic discriminator is constructed. In the **trec** package, we estimated $\boldsymbol{\theta}$ by the maximum likelihood method, and the estimates were obtained by the R function **multinom** of the R package **nnet** (Venables and Ripley, 2002).

Now, we describe how $\mathbf{X}_i$ is defined. Let $N_i$ be the number of time steps for the $i$ th variable. First, we estimate polynomial trends for $m_1$ variables by the method described in the "Trend model" section. Although polynomial trends are estimated by setting time step points as $1,\cdots,N_i$ in the "Trend model" section, here the time step points are standardized to [0, 1], and we define the standardized estimators as $\hat{\gamma}_{i,l}$, $l=0,1,2,3$. Note that the estimated trend components are invariant with respect to standardization of time step points. Let $d_i \in \{1,2,3\}$ be a dimension of the polynomial trend for the $i$ th variable. For example, $\hat{\gamma}_{i,2} = \hat{\gamma}_{i,3} = 0$ when $d_i = 1$. Then, we define $\mathbf{X}_i$ as

$$\mathbf{X}_i \equiv (1, d_{i,1}, d_{i,2}, \hat{\gamma}_{i,0}, \hat{\gamma}_{i,1}, \hat{\gamma}_{i,2}, \hat{\gamma}_{i,3})',$$

where

$$d_{i,1} \equiv \begin{cases} 1 & (d_i = 1) \\ 0 & (d_i \neq 1) \end{cases}, \quad d_{i,2} \equiv \begin{cases} 1 & (d_i = 2) \\ 0 & (d_i \neq 2) \end{cases}.$$

Finally, we explain the procedure used to assign an *icon*. Assume that we have

$$\mathbf{X}_* \equiv (1, d_{*,1}, d_{*,2}, \hat{\gamma}_{*,0}, \hat{\gamma}_{*,1}, \hat{\gamma}_{*,2}, \hat{\gamma}_{*,3})',$$

obtained by estimating the trend for new data. For a linear trend, i.e., $d_{*,1} = 1$, an *icon* is assigned by the slope of the estimated trend. When the new data is classified into the upward group, *icon* 4 is assigned if $\hat{\gamma}_{*,1} > 0$. When the new data is classified into the downward group, *icon* 7 is assigned if $\hat{\gamma}_{*,1} < 0$. When the new data is classified into the flat group, *icon* 1 is assigned if $|\hat{\gamma}_{*,1}| \leq 0.1$. For other cases, the assignment follows the discriminator.

**Using trec**

Time series datasets containing $m$ variables and $N$ time steps should be prepared. As mentioned above, **trec** involves three estimation procedures, which are also conducted interactively for users. The relevant functions and objectives are summarized in Table 1.

| Function | Objective | Implementation by users |
|---|---|---|
| TREC1 | trend estimation | plot of original data<br>plot of each estimated trend<br>plot of all estimated trends |
| TREC2 | common trend classification | set two target trends<br>(default automatically sets two artificial target trends)<br>plot of two/three grouped trends based on output by clustering<br>plot of dendrogram obtained by clustering<br>plot of two/three grouped trends based on discriminant function |
| TREC3 | common trend classification | define multi-group classification<br>show summary table for common trends and *icon* assignment |

Table 1: Functions that users can implement in **trec**

**TREC1**

We apply an example dataset called *exData* in the package, including $m = 9$ variables with $N = 20$ time steps. Here, *exData* takes the form of a "data.frame" object with 20 rows and 10 columns. The first column indicates 'year' corresponding to time points, and it is used for the axis label when describing plots for the original data or trend estimation. An argument of *TREC1* includes a data.frame object with $N$ rows and $m$ columns, like *exData*. *TREC1* estimates trends for each variable, where each variable is standardized and estimated. *TREC1* can be executed as follows:

```
res1 <- TREC1(exData)
```

In *TREC1*, the variable names of *exData* are automatically represented by V1, $\cdots$, V9.

The relationship between the original and the represented names is automatically output on an R console.

If variables are removed for estimation, e.g. there are many missing values at certain time points, they are also displayed under the output:

```
The following variable(s) is/are removed:
```

The missing values between time points (e.g. years) are interpolated. However, if the data show values missing at start and/or end points, the variables are excluded from this analysis. The output of *TREC1* takes the form of a list object and is summarized in Table 2.

| Output | Content |
|---|---|
| fig.RawData | figure of each variable |
| fig.StdData | figure of each standardized variable |
| fig.ctrend | figure of each standardized variable with estimated trend |
| fig.trend | figure of all estimated trends |
| argTREC | list required in later steps<br>    TR: trend estimates<br>    ggD3: a form to draw figures<br>    Y: standardized data<br>    dim: indicator for dimension of polynomial equation<br>    coef: coefficients of polynomial equation |
| remove | removed variables |
| Vnames | original and represented variable names |

Table2: Output of *TREC1*

*TREC1* can output the plots for original data Figure 2, standardized data Figure 3, estimated trend (one figure by gathering all trends) (Figure 4), and estimated trend with 95% prediction bands (Figure 5) as follows:

```
res1$fig.RawData
res1$fig.StdData
res1$fig.trend
```

and

```
res1$fig.ctrend .
```

If $m_0 > 16$, *fig.ctrend* takes the form of a list object, where $m_0 \, (\leq m)$ is the number of variables subjected to estimation. Hence, the above code is replaced by

```
plot(res1$fig.ctrend[[1]]); plot(res1$fig.ctrend[[2]]); ...
```

*argTREC* takes the form of a list object, which includes *TR*, *ggD3*, *Y*, *dim*, and *coef*. *TR* takes the form of a matrix object with $N$ rows and $m_0$ columns, which includes prediction by the trend model for each variable. *ggD3* takes the form of a data.frame object with $m_0 N$ rows and 4 columns, which is used to draw figures in later steps. *Y* takes the form of a matrix object with $N$ rows and $m_0$ columns, which has standardized data for each variable. *dim* and *coef* take the form of matrix objects with $m_0$ rows and 2 and 4 columns, which are defined by

$$dim:\begin{pmatrix} d_{1,1} & d_{1,2} \\ \vdots & \vdots \\ d_{m_0,1} & d_{m_0,2} \end{pmatrix}, \quad coef:\begin{pmatrix} \hat{\gamma}_{1,0} & \cdots & \hat{\gamma}_{1,3} \\ & \vdots & \\ \hat{\gamma}_{m_0,0} & \cdots & \hat{\gamma}_{m_0,3} \end{pmatrix}.$$

**TREC2**

This step performs rough classification of the estimated trends into two (upward and downward) or three (upward, flat, and downward) groups based on the two target trends $\hat{T}_1(n)$ and $\hat{T}_2(n)$. In the default setting, these two trends are fixed as linear trends. Using the output for *argTREC*, *TREC2* can be executed as follows:

```
argTREC <- res1$argTREC
res2 <- TREC2(argTREC)
```

If users want to select target trends from the estimated trends, they can input two variables in *pvar* as

```
res2 <- TREC2(argTREC, pvar=c("V2", "V7"))
```

Moreover, two-category discrimination can be expanded to three-category discrimination by the following option:

```
res2 <- TREC2(argTREC, groups=3)
```

In Solvang and Planque (2020), hierarchical clustering was applied using the two-category discriminant function as a distance measurement. Two or three groups were detected by the dendrogram with a centroid link of clusters. This flow was obtained according to the method in Solvang et al. (2008), and *TREC2* calculated a two-category discriminant function and made a dendrogram based on the following function:

```
res2c_2g <- TREC2(res1$argTREC) # two groups detection by clustering based
                                # on two-category discriminant function
res2c_2g$dend() # show the dendrogram
res2c_3g <- TREC2(res1$argTREC, group = 3)# three groups detection by
                                           # clustering based on two-category
                                           # discriminant function
res2c_3g$dend() # show the dendrogram
```

The trend groups detected by clustering based on two-category discrimination are presented in Figure 6. The dendrograms corresponding to these groups are shown in Figure 7. The panel on the right-hand side of Figure 7 shows that one more group is divided into upward and flat trends in the estimated linear trends (V2, 8, 5 and 7). The difference between two and three categories is whether the estimated trend for V2 belongs to upward or flat. As seen in Figure 5, if it is meaningful for a user to detect the trend of V2 as flat compared with the other linear trends, it should belong to the flat group. Alternatively, we can also detect roughly common trend groups without clustering as follows:

```
res2d <- TREC2(res1$argTREC, clustering=FALSE)
```

The output for trend groups is shown in Figure 8. This procedure indicates that variable 2 belonged to the downward group. Even if we perform the command

```
res2d_3g <- TREC2(res1$argTREC, clustering=FALSE, group=3),
```

*TREC2* returns the following message to us:

```
The following group(s) is/are not applicable:
[1] "Flat",
```

which means that three-category discrimination is not applicable in this case. As seen in Figure 8, the trend of V2 belonging to the downward group may be useful when the intended classification is limited to either upward or downward discrimination. On the other hand, the trend of V2 could also lead to belonging to the flat group as shown above. The **trec** in R includes two options, detecting groups by clustering based on two-category discrimination and by discrimination alone. The final decision for the outputs obtained by those options would depend on the aim of the user. Comparing detected common trend groups with the estimated trend pattern summarized in Figure 5 is an important step in interpreting the best grouping of common trends for the user.

### TREC3

This step performs multi-category discrimination to classify more common trend groups based on target trends selected by the user, and it assigns *icon*s to common trend groups. If we set the target trends for V1, 6, and 9 to downward, V8 to upward, and V2 to flat, the following setting is required:

```
tvard <- list(
Downward = paste0("V", c(1,6,9)),
Upward = paste0("V", c(8)),
Flat = paste0("V", c(2))
)
```

Using this, *TREC3* can be executed as follows:

```
res3_2c3g <- TREC3(tvard, res2c_3g$trn, res1$argTREC)
```

*TREC3* outputs a table of assigned *icon*s and variable names for each group (Figure 9). The output by *TREC3* can show trend groups for downward, upward, and flat separately as summarized in Table 3.

| Output | Content |
|---|---|
| fig.down | trend groups based on selected target trends in downward |
| fig.up | trend groups based on selected target trends in upward |
| fig.flat | trend groups based on selected target trends in flat |
| fig.icon | figure of assigned *icon*s |

Table 3: Contents of the output obtained by *TREC3*

The *icon*s are automatically assigned based on the information by learning many trend patterns. For the predefined *icon*s in Figure 1, the **trec** in R was set with the rules that *icon*s 2, 7, and 8 be assigned to the downward group, *icon*s 3, 4, and 5 be assigned to the upward group, and *icon*s 1, 6, and 9 be assigned to the flat group. For example, the target trend of V6 was interpreted by the **trec** in R as any *icon* being available for assignment. However, if we review the trends of V3, 4, and 6 in Figure 5, the patterns may be recognized as *icon* 8. Therefore, the output for *icon* assignment should be confirmed by comparison with the estimated trend according to the aim of the user.

## Conclusion

We presented the R package **trec** for analysing common trends in a marine ecosystem by using statistical trend estimation and classification. The original idea of TREC has been applied annually to the ICES integrated assessment working groups since 2019. Developing the R package **trec** is expected to contribute to the IEAtools package (Otto, 2022) and to help research professionals to more widely and practically conduct trend analysis in integrated ecosystem assessment (ICES, 2018).

## Acknowledgement

This work was supported by JSPS Bilateral Program Grant Number JPJSBP120219927 and the second author's research was partially supported by JSPS KAKENHI Grant Numbers JP20H04151 and JP21K13834.

## Reference

Akaike,H. A new look at the statistical model identification. *IEEE Transactions on Automatic Control*, 19:716--723, 1974. https://doi.org/10.1109/TAC.1974.1100705.

Amri, M., Sumertajaya, I., and Syafitri, U. Comparison of multinomial logistic discriminant analysis (mlgda) and classification and regression tree (cart) performance in classifying the impact of working children. *Journal of Physics: Conference Series*, 1490(1):012033, 2020. https://doi.org/10.1088/1742-6596/1490/1/012033.

ICES. Report of the Workshop on Integrated Trend Analyses in Support to Integrated Ecosystem Assessment (WKINTRA). *ICES Document CM*, 15:23pp, 2018.

ICES. Final report of the working group on the integrated assessment of the Norwegian Sea (WGINOR). *ICES Document CM*, 10:149pp, 2020.

Jeune, W., Francelino, M., de Souza, E., Filho, E., and Rocha, C. Multinomial logistic regression and random forest classifiers in digital mapping of soil classes in western Haiti. *Revista Brasileira de Ciência do Solo*, 42:e0170133, 2018. https://doi.org/10.1590/18069657rbcs20170133.

Levin, P.S. and Fogarty, M.K. and Murawski, S.A. and Fluharty, D. Integrated ecosystem assessments: developing the scientific basis for ecosystem-based management of the ocean. *PLoS Biology*, 7(e14):299-314, 2009.

Mirzaei, S. Hyperspectral image classification using non-negative tensor factorization and multinomial logistic regression. *Journal of Applied Remote Sensing*, 13(2): 1-18, 2019. https://doi.org/10.1117/1.JRS.13.026501.

Otto, S. *Integrated Ecosystem Assessment tools*, ICES WKINTRA, Humberg, Germany, 2022.

https://gitlab.rrz.uni-hamburg.de/saskiaotto/IEAtools.

R Core Team. R: A Language and Environment for Statistical Computing. R Foundation for Statistical Computing, Vienna, Austria, 2016. https://www.R-project.org/. ISBN 3-900051-07-0.

Solvang, H.K. and Planque, B., Estimation and classification of temporal trends to support integrated ecosystem assessment. *ICES Journal of Marine Science*, 77(7-8): 2529-2540. https://doi.org/10.1093/icesjms/fsaa111.

Solvang, H.K., Taniguchi, M., Nakatani, T. and Amano, S. Classification and similarity analysis of fundamental frequency patterns in infant spoken language acquisition. *Statistical Methodology*, 5(5): 187-208, 2008. https://doi.org/10.1016/j.stamet.2007.08.005.

Venables, W.N. and Ripley, B.D. *Modern Applied Statistics with S*. Springer, New York, fourth edition, 2002. https://www.stats.ox.ac.uk/pub/MASS4. ISBN 0-387-95457-0.

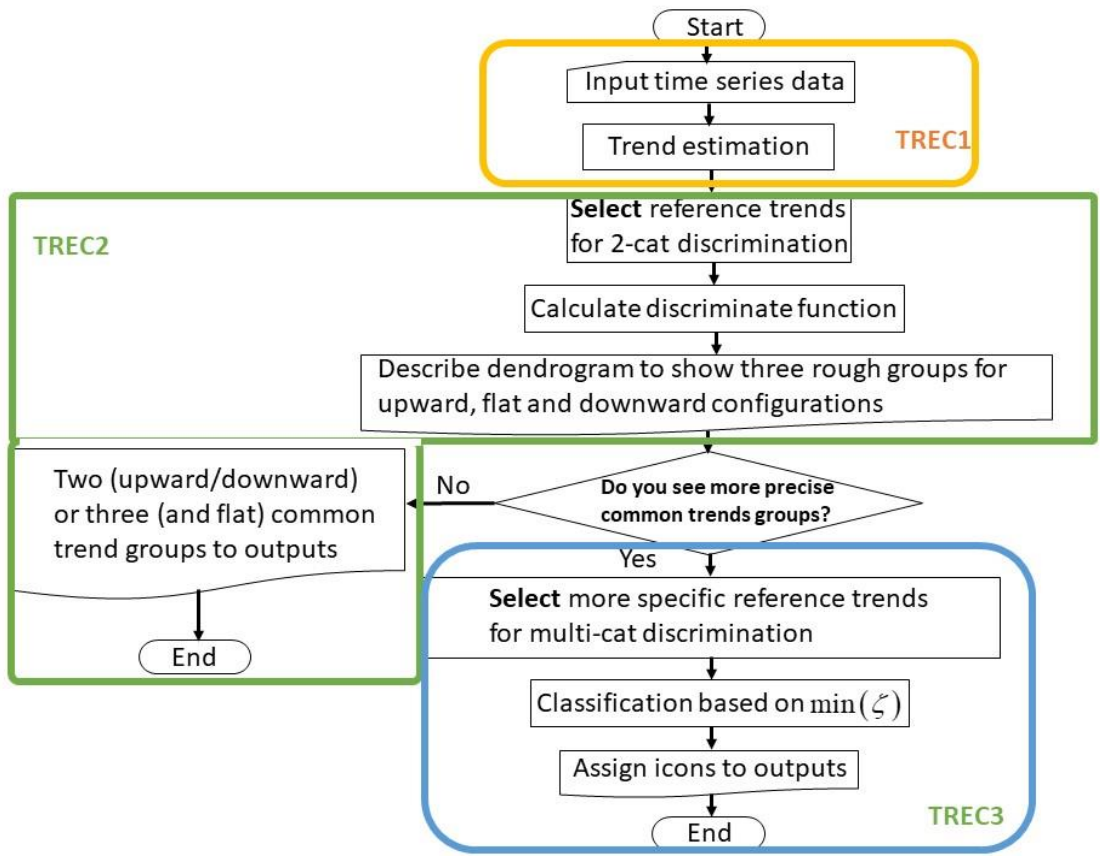

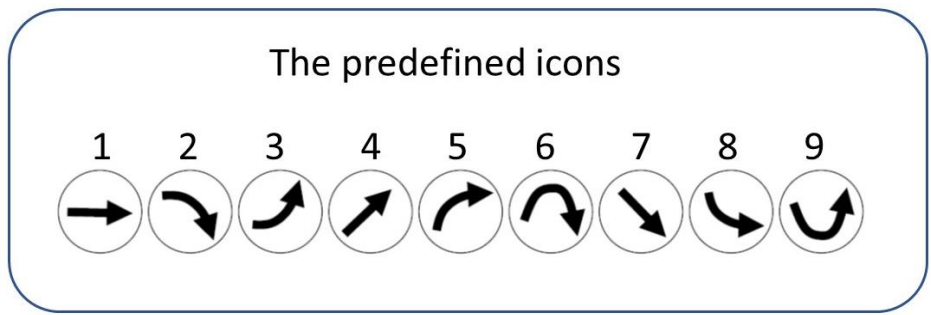

Figure 1: Flow chart of procedure in **trec** and the predefined *icon*s used in TREC3

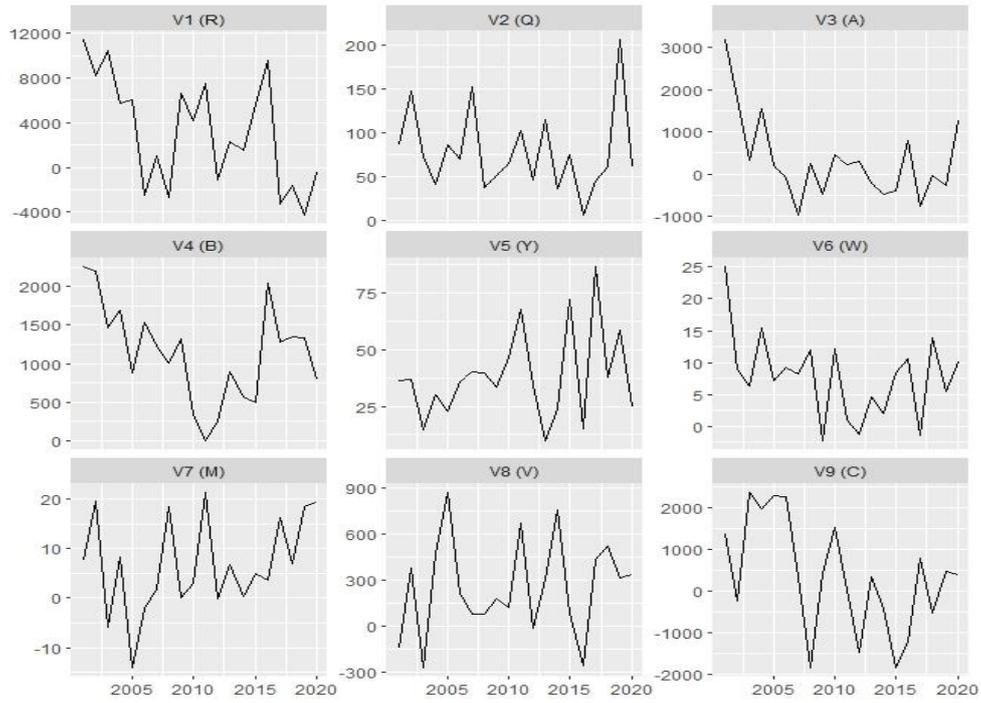

Figure 2: Output by fig.RawData

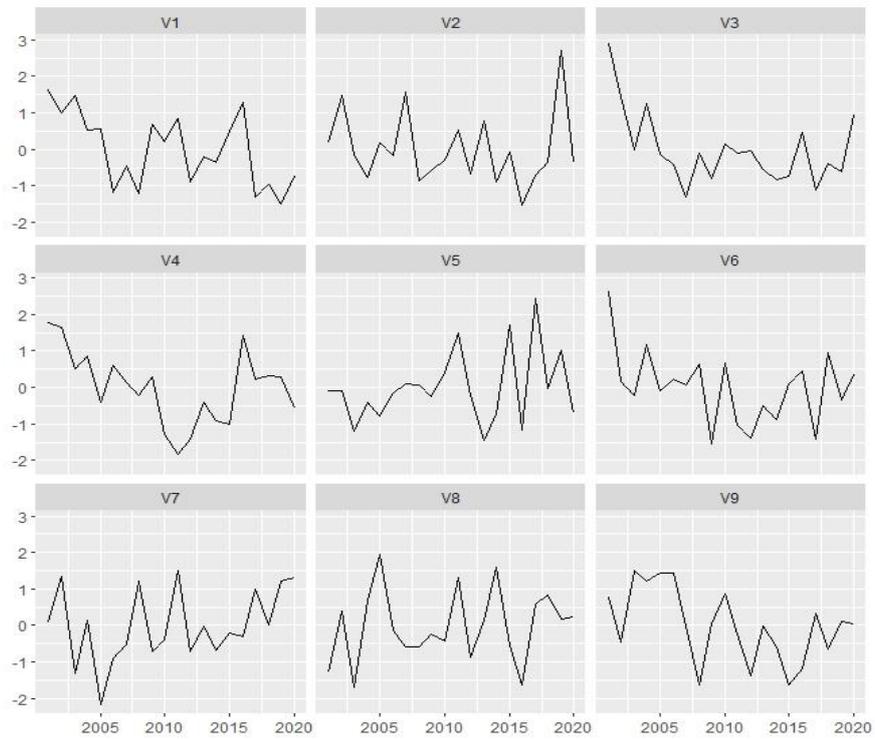

Figure 3: Output by fig.StdData

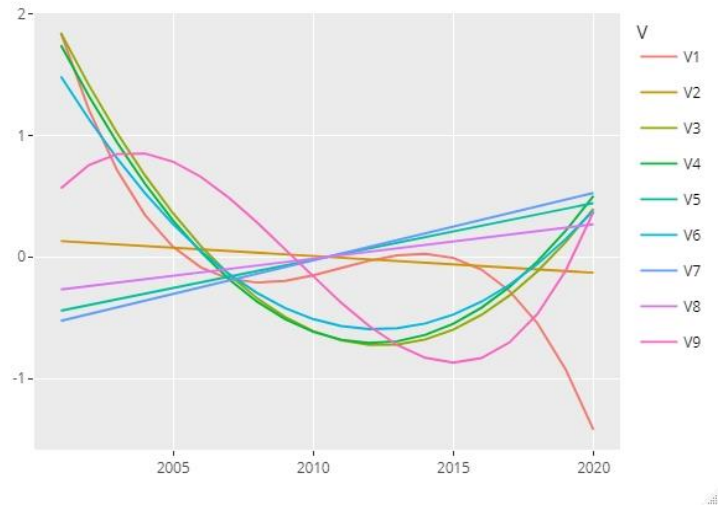

Figure 4: Output by fig.trend

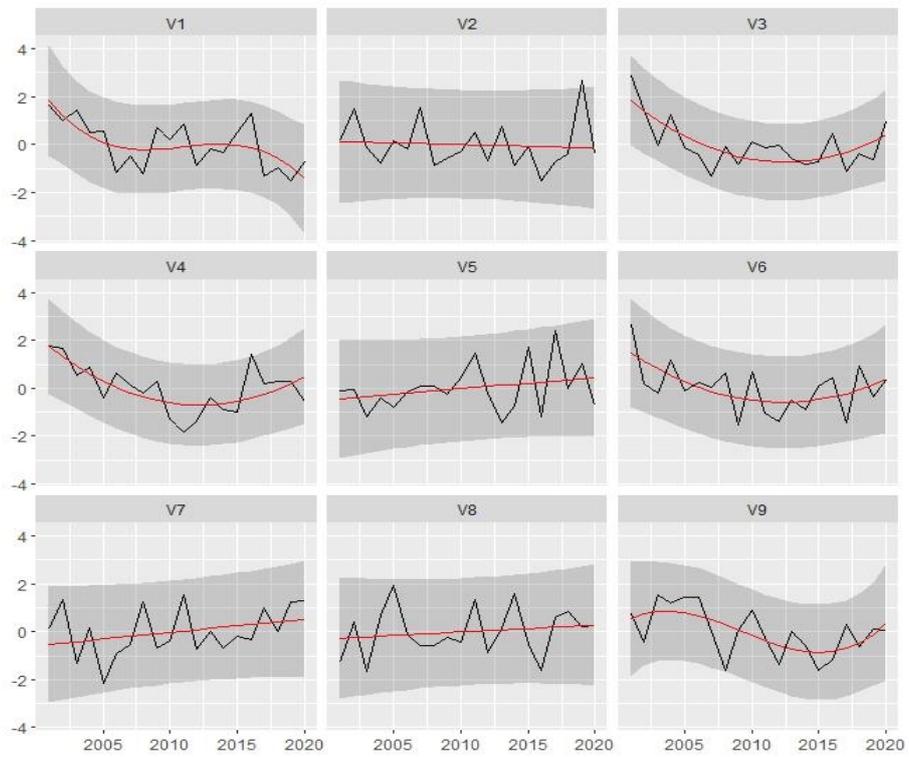

Figure 5: Output by fig.ctrend. Red and black lines indicate standardized data and estimated trend. The grey-coloured area corresponds to a 95% prediction band for the trend.

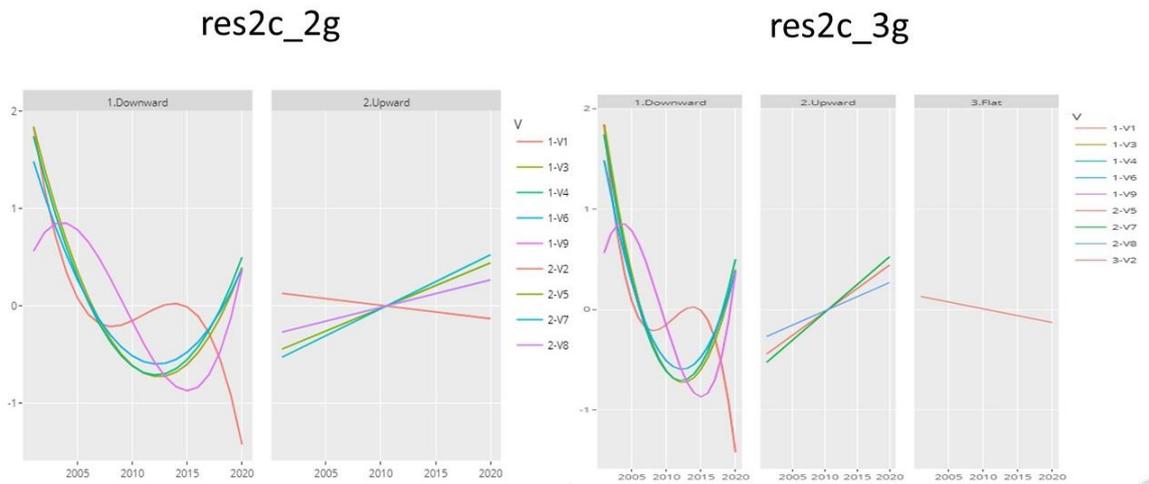

Figure 6: Trend groups obtained using two- or three-group detection by clustering based on two-category discrimination.

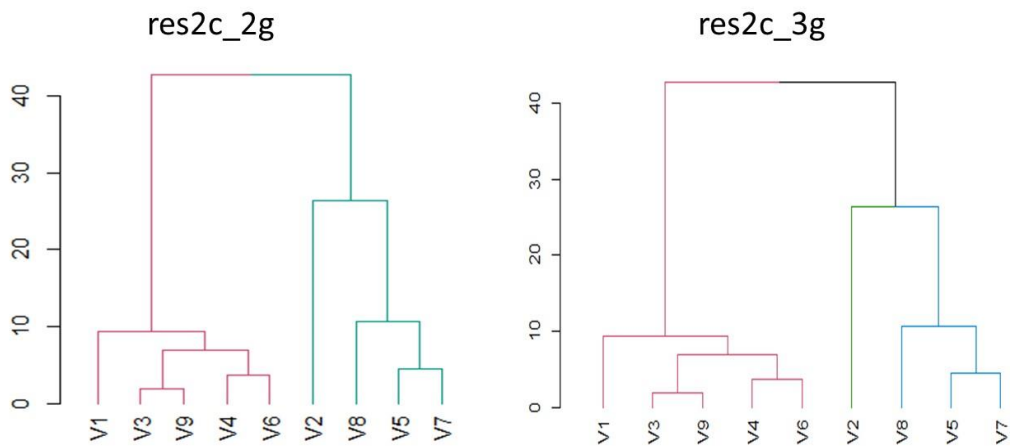

Figure 7: Dendrograms obtained by clustering with centroid link and distance obtained by the discriminant function.

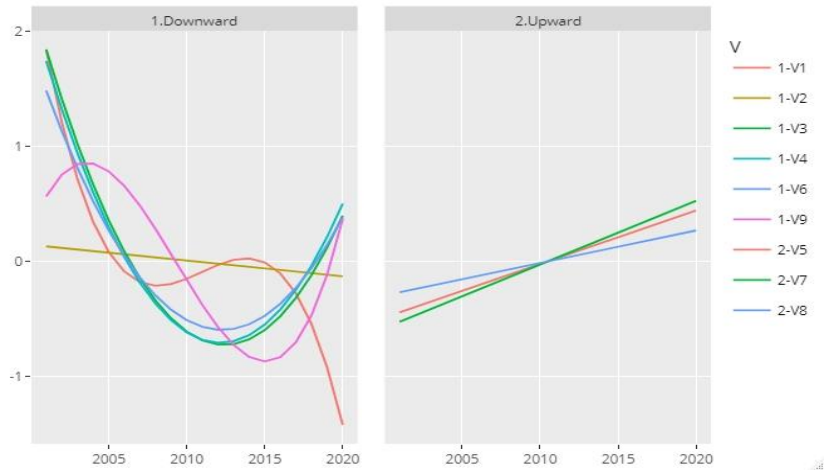

Figure 8: Trend groups obtained by two-category discriminant analysis without clustering.

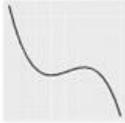

Figure 9: Outputs by TREC3